\documentclass[
 aip,
 apl,
 amsmath,amssymb,
preprint,%
]{revtex4-1}

\usepackage{graphicx}% Include figure files
\usepackage{dcolumn}% Align table columns on decimal point
\usepackage{bm}% bold math
\usepackage{color,soul}
\usepackage[utf8]{inputenc}
\usepackage[T1]{fontenc}
\usepackage{mathptmx}
\usepackage{etoolbox}

\makeatletter
\def\@email#1#2{%
 \endgroup
 \patchcmd{\titleblock@produce}
  {\frontmatter@RRAPformat}
  {\frontmatter@RRAPformat{\produce@RRAP{*#1\href{mailto:#2}{#2}}}\frontmatter@RRAPformat}
  {}{}
}%
\makeatother
\begin{document}

\title{Prominent phonon transmission across aperiodic superlattice through coherent mode-conversion}
\author{Theodore Maranets}
\author{Yan Wang}%
\affiliation{Department of Mechanical Engineering, University of Nevada, Reno, Reno, NV, 89557, USA }%
\email{tmaranets@unr.edu}

\begin{abstract}
In both particle and wave descriptions of phonons, the dense, aperiodically arranged interfaces in aperiodic superlattices are expected to strongly attenuate thermal transport due to phonon-interface scattering or broken long-range coherence. However, considerable thermal conductivity is still observed in these structures. In this study, we reveal that incoherent modes propagating in the aperiodic superlattice can convert, through interference, into coherent modes defined by an approximate dispersion relation. This conversion leads to high transmission across the aperiodic superlattice structure, which contains hundreds of interfaces, ultimately resulting in significant thermal conductivity. Such incoherent-to-coherent mode conversion behavior is extensively observed in periodic superlattices. This work suggests an effective strategy to manipulate the phonon band structure through layer patterning or material choice, enabling precise control of phonon transmission across aperiodic superlattices.
\end{abstract}

\maketitle

Superlattices (SLs), which are metamaterials composed of alternating layers of two or more materials, exhibit secondary periodicity that results in unique thermophysical properties \cite{xie2018phonon,anufriev2021review,zhang2021coherent,qian2021phonon}. These properties enable various novel applications, including quantum cascade lasers and thermoelectric materials \cite{qian2021phonon,shi2015evaluating,volz2016nanophononics}. The densely packed interfaces within SLs can induce significant phonon-interface scattering. Moreover, strong interference of scattered phonons manifest as wave-like phonons that propagate through the interfaces coherently without scattering \cite{maldovan2015phonon,anufriev2021review,zhang2021coherent}. A rigorous understanding of coherent phonon transport in SLs is of both fundamental significance for comprehending phonon behavior in complex materials and practical importance for thermal management in modern devices utilizing SL structures. 

Recently, the disruption of the secondary periodicity in SLs has garnered significant attention due to its effectiveness in substantially reducing the lattice thermal conductivity of otherwise periodic SLs \cite{wang2014decomposition,wang2015optimization,chakraborty2020quenching,chakraborty2020complex}. Disrupted phonon coherence and destructive interference, particularly Anderson localization, are hypothesized to be the primary mechanisms, supported by extensive yet mostly indirect evidence from spectral phonon analysis and length-dependent thermal conductivity data obtained from atomistic simulations \cite{wang2014decomposition,wang2015optimization,ma2020dimensionality,hu2021direct}. Furthermore, it has been demonstrated that the lattice thermal conductivity of well-randomized SLs can be lower than that of random alloys \cite{wang2015optimization,chakraborty2017ultralow}. Ultimately, achieving full-spectrum suppression of phonon transmission across aperiodic SLs to attain the lowest possible lattice thermal conductivity is desirable. Recent studies employing machine learning have optimized the layer thicknesses of aperiodic SLs to minimize thermal conductivity, though the precise phonon mechanisms responsible for this minimized conductivity remain elusive \cite{chakraborty2020quenching,chowdhury2020machine}.

In this study, we aim to address two fundamental questions regarding phonon transport in aperiodic SLs, which are characterized by aperiodically arranged interfaces between two distinct materials. First, we will examine the behavior of incoherent phonon modes that adhere to the phonon dispersion relations of the SL's base material. It is anticipated that these modes would experience scattering at the densely packed interfaces within the aperiodic SL, leading to diminished transmission across the device. However, the validity of this assumption remains to be thoroughly investigated. Both limited experimental data and modeling suggest that aperiodic SLs, even those optimized through machine learning algorithms, still exhibit considerable thermal conductivity, which suggests the existence of extended phonon modes that can readily transport across the aperiodic device \cite{wang2014decomposition,wang2015optimization,chakraborty2020quenching,chakraborty2020complex,chowdhury2020machine}. Second, we seek to determine whether incoherent modes retain their original form or undergo conversion into new modes corresponding to the aperiodic SL structure as a result of complex interference events at the interfaces. While previous studies have hypothesized such mode-conversion, direct evidence supporting this hypothesis is currently lacking \cite{tamura1990acoustic,schelling2003multiscale,jiang2021total,maranets2024influence}.

\begin{figure}
    \centering
\includegraphics[width=0.8\textwidth]{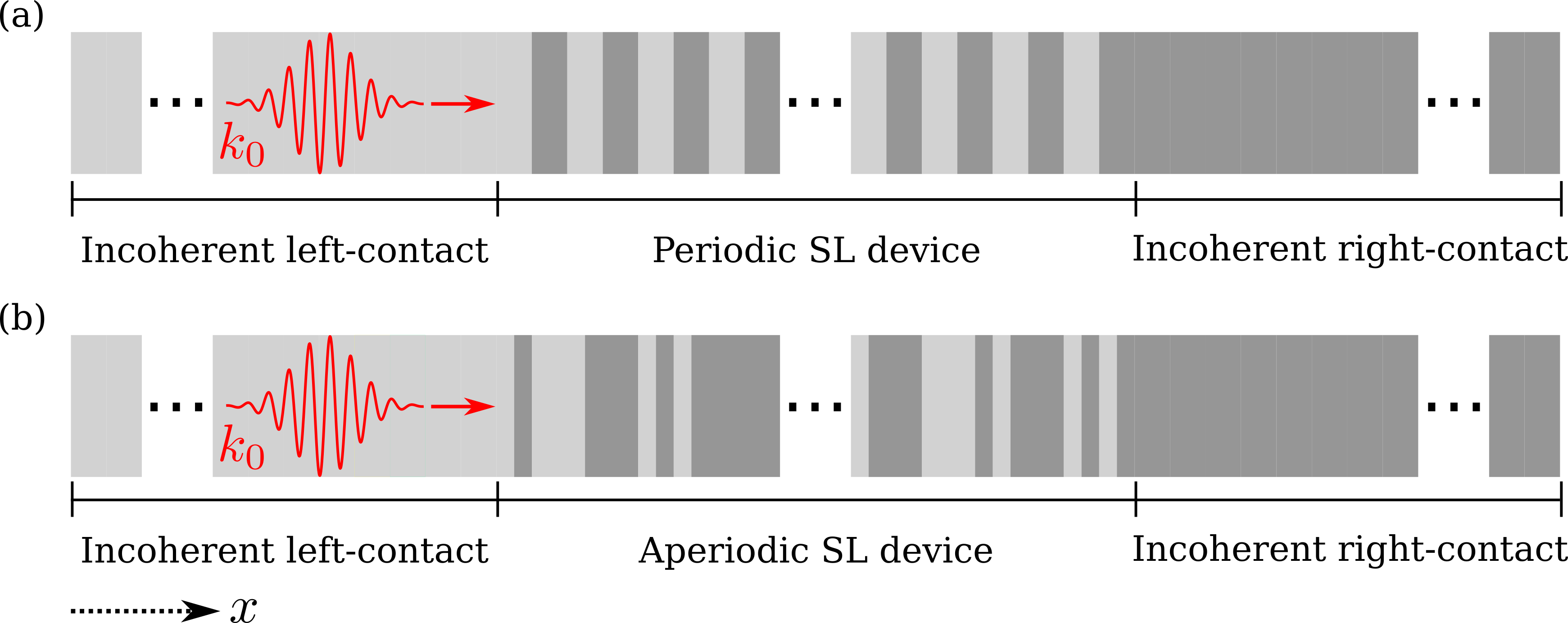}
    \caption{Schematic illustrations of the simulation domain with periodic (a) and aperiodic SL (b) devices. The incident LA-mode incoherent phonon wave-packet centered at wavevector $k_{0}$ in reciprocal-space is generated in the left-contact and allowed to propagate into the device. The total energies of the left- and right-contacts, as well as the device are monitored throughout the simulation to compute the transmission across the device. Note: the illustrated sizes of the wave-packet and wavelength are not to scale.}
    \label{fig:figsysdiagram}
\end{figure}

Atomistic phonon wave-packet simulations provide a direct means of analyzing phonon wave dynamics, with the ability to resolve mode, wavevector, and spatial coherence length \cite{schelling2002phonon, schelling2003multiscale,schelling2004kapitza}. This allows for a rigorous examination of phonon scattering by specific heterogeneous structures, making the method particularly useful for investigating fundamental transport physics in metamaterials with complex structures. The technique employed in this study is adapted from the method developed by Schelling et al. \cite{schelling2002phonon}. We utilize the same system configuration as in our previous work \cite{maranets2024influence} as the model device.
However, distinct from Ref.~\citenum{maranets2024influence}, in this study, we simulate an incoherent longitudinal-acoustic (LA) mode derived from the dispersion relation of one of the constituent materials (Fig. S1a) of the SL, as opposed to a coherent mode derived from the SL's dispersion relation (Fig. S1b) as used in Ref.~\citenum{maranets2024influence}. Consequently, the left- and right-contacts in the simulation domain consist of two pure materials in this study, in contrast to the periodic SL contacts used in Ref.~\citenum{maranets2024influence}. Additionally, unlike coherent modes, which can transmit across their corresponding SL devices without scattering, the incoherent phonon modes studied here exhibit significant scattering at the interfaces of both periodic and aperiodic SL devices (Figs.~\ref{fig:figsysdiagram}a and~\ref{fig:figsysdiagram}b, respectively), resulting in non-unity transmission.

A phonon wave-packet is described by the following equation for atom displacement:
\begin{eqnarray}
u_{i,n} = \frac{A_{i}}{\sqrt{m_{i}}}\varepsilon_{k_{0},i}\exp{(i[k_{0}\cdot(x_{n} - x_{0})-\omega_{0}t]})\exp{(-4(x_{n}-x_{0}-v_{g0}t)^{2}/l_{c}^{2})}\qquad 
\label{eqn:pwp}
\end{eqnarray}
where $\varepsilon_{k_{0},i}$, $\omega_{0}$, and $v_{g0}$ are the associated eigenvector, frequency, and group velocity, respectively for the phonon centered at wavevector $k_{0}$ in reciprocal-space.  $A_{i}$ is the wave amplitude, $m_{i}$ is the mass of atom $i$, $x_{n}$ is the position of the $n$th unit cell containing atom $i$, $x_{0}$ is the initial position of the wave-packet, and $l_{c}$ is the spatial coherence length. The wave-packet is generated by setting the real parts of Eqn.~\ref{eqn:pwp} and the time-derivative of Eqn.~\ref{eqn:pwp} evaluated at time $t=0$ as the initial atom displacement and velocity, respectively. Since a large coherence length aids the observation of wave interference effects  \cite{zhang2021coherent,latour2014microscopic,latour2017distinguishing, maranets2024influence}, we set the spatial coherence length to be four times the device length in all simulations conducted in this study. Other parameters such as the boundary conditions, size of the device, cross-sectional area, and transmission calculation are the same as Ref.~\citenum{maranets2024influence}. The simulations are performed using the LAMMPS package \cite{thompson2022lammps}.

\begin{figure}
    \centering
    \includegraphics[width=\textwidth]{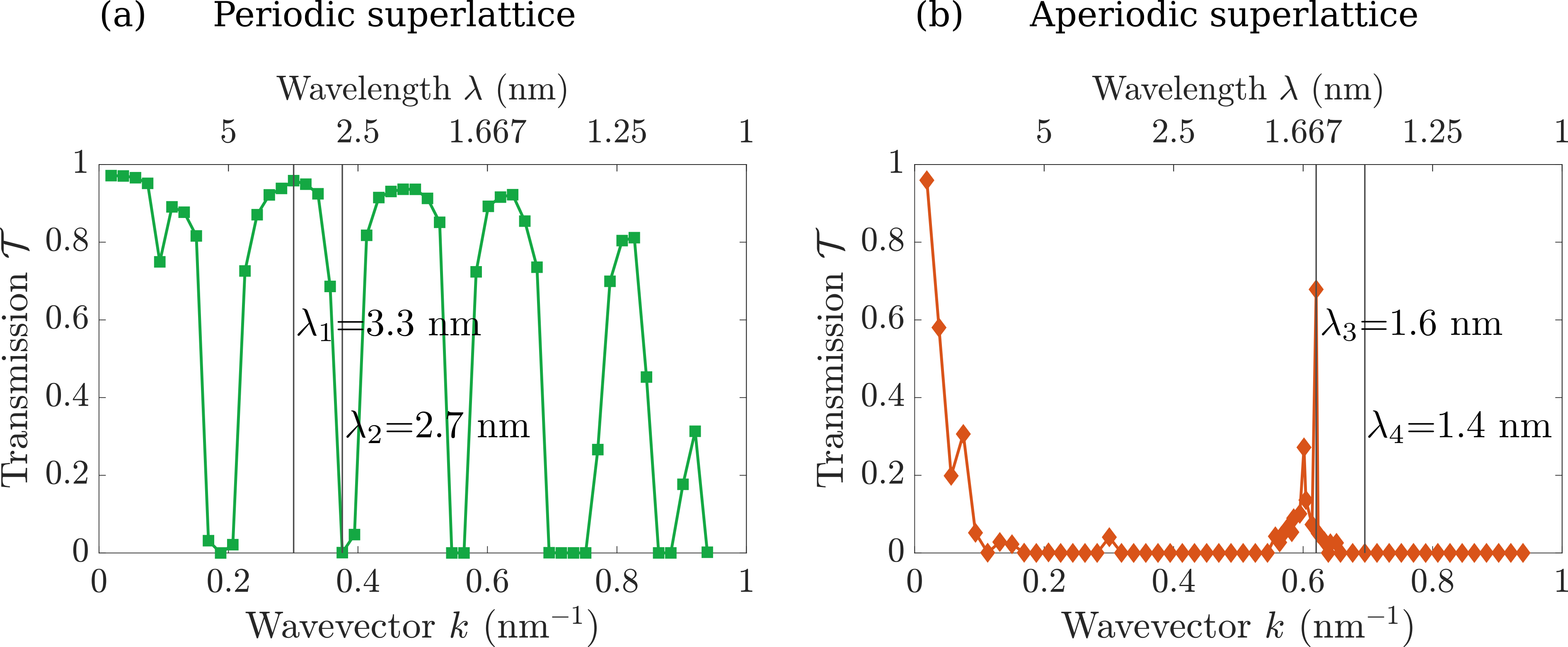}
    \caption{Transmission $\mathcal{T}$ versus wavevector $k$ (or inverse of wavelength $\lambda$) for the LA-mode incoherent phonon wave-packet propagating through the periodic (a) and aperiodic SL (b) devices as illustrated in Figs.~\ref{fig:figsysdiagram}a and~\ref{fig:figsysdiagram}b, respectively. Key wavelengths $\lambda_{1}=3.3$ nm, $\lambda_{2}=2.7$ nm, $\lambda_{3}=1.6$ nm, and $\lambda_{4}=1.4$ nm analyzed in this letter are indicated in the figures by the vertical black lines. Only the frequencies below the cutoff frequency of the two constituent layer materials of the SL are studied.}
    \label{fig:transmission}
\end{figure}

In Figs.~\ref{fig:transmission}a and~\ref{fig:transmission}b, we present the transmission of the LA-mode incoherent phonon wave packet through the periodic and aperiodic SL devices, respectively. The near-unity phonon transmission in the long-wavelength (or small wavevector) limit of both figures is due to the wavelengths being significantly larger than the period thicknesses, and thus minimally affected by the interfaces\cite{latour2014microscopic,seyf2016method,latour2017distinguishing,maranets2024influence}. The periodic SL spectrum exhibits an oscillatory pattern of near-unity and zero transmission values, with the highest transmission occurring at the longest wavelengths and a slightly decreasing trend as the wavelength shortens. In contrast, the aperiodic SL spectrum shows strong transmission at the long-wavelength limit, while the remainder of the spectrum is flattened at zero, except for a narrow but prominent transmission peak at an intermediate wavelength. 

We emphasize the prominent transmission of intermediate-wavelength incoherent phonons across the aperiodic SL (Fig.~\ref{fig:transmission}b), which, to our knowledge, has not been previously reported and is quite surprising, regardless of whether the phonons are treated as particle-like or wave-like. If phonons are particle-like, they are expected to be severely scattered by the 256 densely packed interfaces (with an average separation of 1 nm) in the aperiodic SL. Conversely, if phonons are wave-like, the aperiodic (or disordered) SL structure is not expected to support any long-range, extended phonon modes. In both scenarios, we would anticipate nearly zero transmission of incoherent phonons across the aperiodic SL.

To elucidate the prominent transmission peak in the aperiodic SL case, we first focus on understanding the transmission values in the periodic SL. The locations of the transmission dips in Fig.~\ref{fig:transmission}a closely match previous first-principles analyses of Bragg reflection of phonons in the cross-plane direction of the periodic SL \cite{tamura1988acoustic, tamura1988acousticmulti, tamura1989localized} as well as phonon-imaging experiments \cite{hurley1987imaging,tanaka1998phonon}. When the Bragg scattering condition is met, the interference of phonon waves results in total reflection, attenuating energy transport and producing band gaps in the center and edges of the periodic SL Brillouin zone. Conversely, the high transmission of incoherent modes in a periodic SL is suggested to arise from mode-conversion (through interference) to coherent modes, which follow the periodic SL dispersion. Otherwise, the strong interface scattering of incoherent modes by the multiple layers in the device would result in nearly zero transmission across the SL.

\begin{figure}
    \centering
    \includegraphics[width=\textwidth]{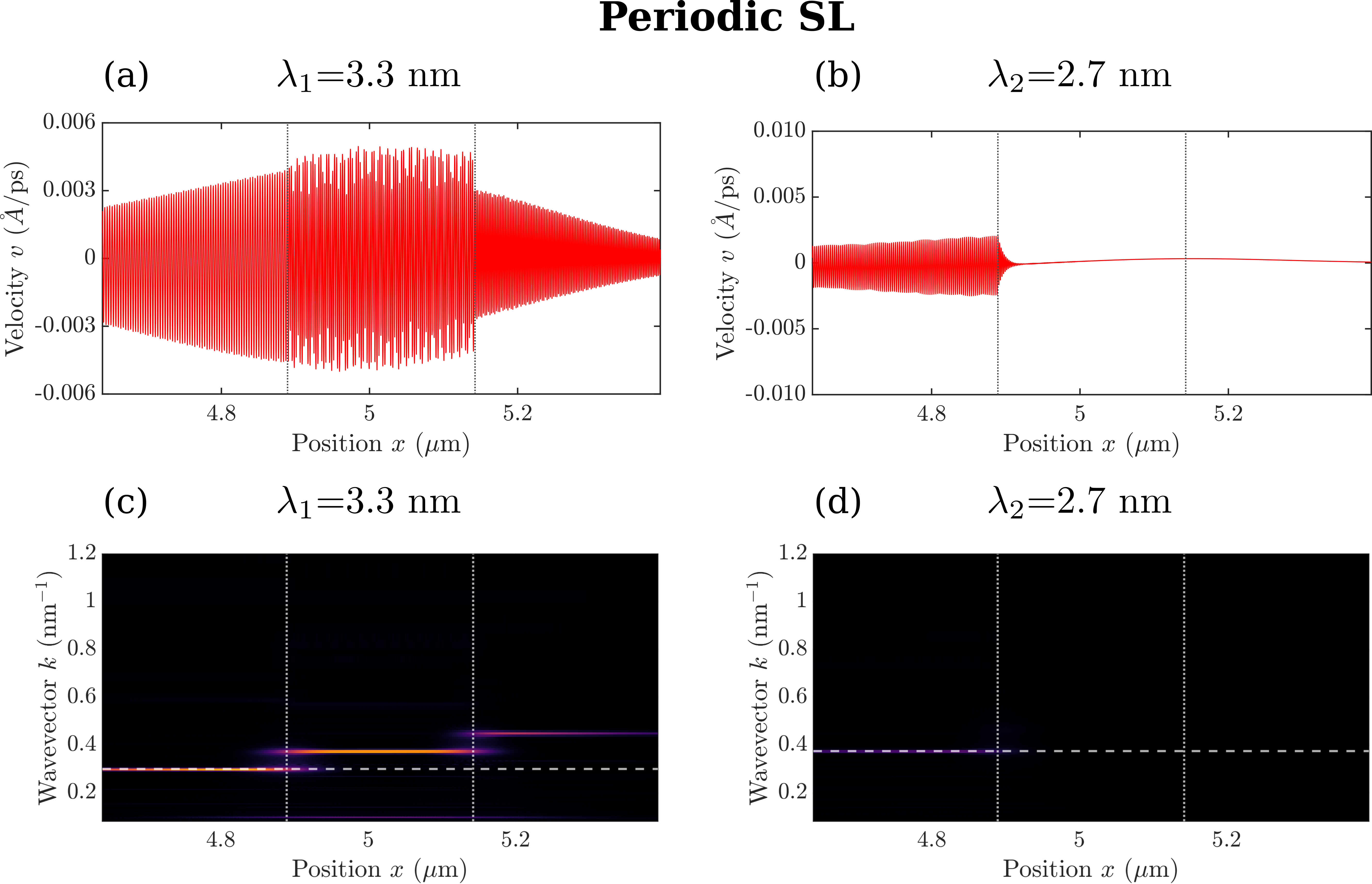}
    \caption{Snapshots of the atom velocity versus real-space position when the wave-packet scatters through the periodic SL device for wavelengths $\lambda_{1}=3.3$ nm (a) and $\lambda_{2}=2.7$ nm (b), representative cases of near-unity and zero transmission as shown in Fig.~\ref{fig:transmission}a, respectively. Snapshots of the reciprocal-space wavelet transform when the wave-packet scatters through the periodic SL device for wavelengths $\lambda_{1}$ (c) and $\lambda_{2}$ (d). The dotted vertical lines denote the position and thickness of the device. The dashed horizontal line denotes the central wavevector $k_{0}$ of the incident wave-packet. Illuminated regions in the heat map not centered about the $k_{0}$ line indicate mode-conversion.}
    \label{fig:SLvelwave}
\end{figure}

To elucidate the mechanisms of transmission, we conduct several quantitative analyses of the wave dynamics. In Fig.~\ref{fig:SLvelwave}, we present the plots of atom velocity versus real-space position and the reciprocal-space wavelet transform \cite{baker2012application} for two wavelengths marked in Fig.~\ref{fig:transmission}a, $\lambda_{1}$ and $\lambda_{2}$, representative cases of near-unity and zero transmission in the periodic SL, respectively. Examining Figs.~\ref{fig:SLvelwave}a and~\ref{fig:SLvelwave}c, we find that high transmission is associated with the incident incoherent phonon wave-packet mode-converting inside the periodic SL device. Similarly, the zero transmission of $\lambda_{2}$ is owed to the incoherent mode not mode-converting as seen in Figs.~\ref{fig:SLvelwave}b and~\ref{fig:SLvelwave}d.

\begin{figure}
    \centering
    \includegraphics[width=\textwidth]{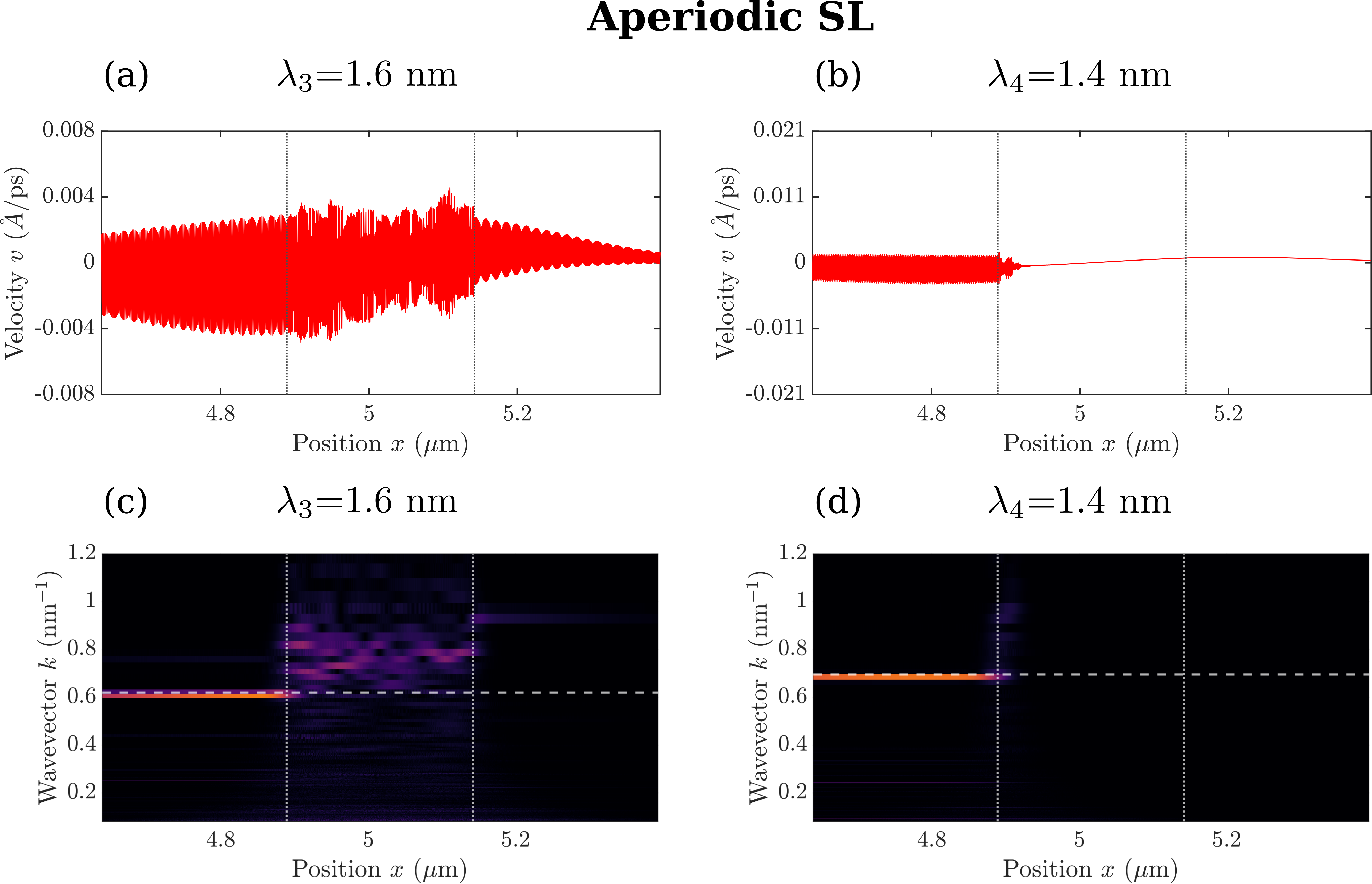}
    \caption{Snapshots of the atom velocity versus real-space position when the wave-packet scatters through the aperiodic SL device for wavelengths $\lambda_{3}=1.6$ nm (a) and $\lambda_{4}=1.4$ nm (b), representative cases of the transmission peak and zero transmission as shown in Fig.~\ref{fig:transmission}b, respectively. Snapshots of the reciprocal-space wavelet transform when the wave-packet scatters through the aperiodic SL device for wavelengths $\lambda_{3}$ (c) and $\lambda_{4}$ (d). The dotted vertical lines denote the position and thickness of the device. The dashed horizontal line denotes the central wavevector $k_{0}$ of the incident wave-packet. Illuminated regions in the heat map not centered about the $k_{0}$ line indicate mode-conversion.}
    \label{fig:RMLvelwave}
\end{figure}

We repeat these analyses for the aperiodic SL, focusing on two wavelengths marked in Fig.~\ref{fig:transmission}b: $\lambda_{3}$ (transmission peak) and $\lambda_{4}$ (zero transmission). Studying the velocity plots and wavelet transforms of the two wavelengths in Fig.~\ref{fig:RMLvelwave}, we observe that the prominent transmission of $\lambda_{3}$ (Figs.~\ref{fig:RMLvelwave}a and~\ref{fig:RMLvelwave}c) is a consequence of substantial mode-conversion. Likewise, the $\lambda_{4}$ phonon does not undergo mode-conversion, as shown in Figs.~\ref{fig:RMLvelwave}b and~\ref{fig:RMLvelwave}d. 

\begin{figure}
    \centering
    \includegraphics[width=\textwidth]{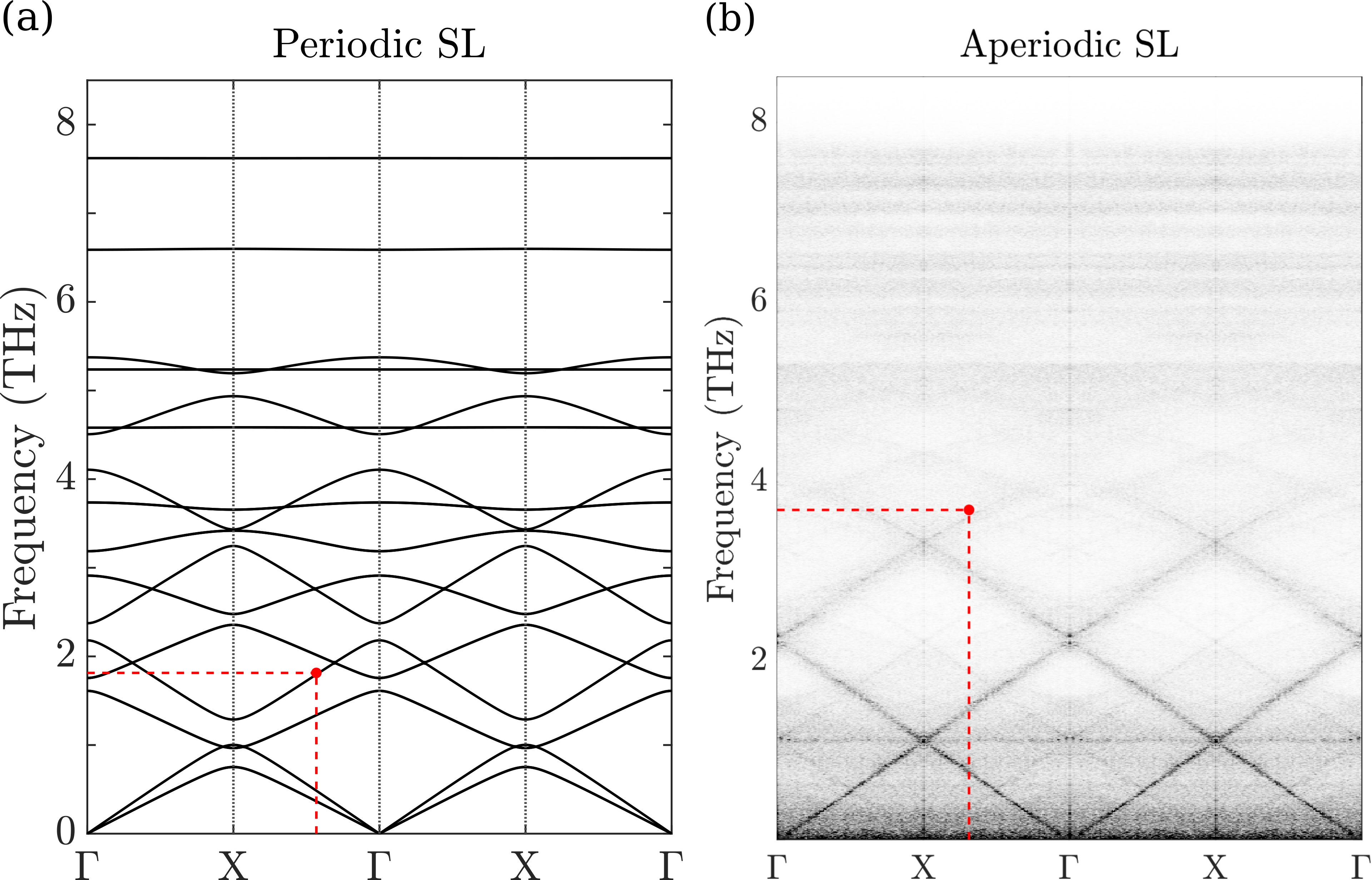}
    \caption{(a) Phonon dispersion relation of the periodic SL with the mode-converted phonon in the $\lambda_{1}=3.3$ nm case (Fig.~\ref{fig:SLvelwave}c) located with the red dot and dashed lines. (b) Spectral energy density heat map of the aperiodic SL with the mode-converted phonon in the $\lambda_{3}=1.6$ nm case (Fig.~\ref{fig:RMLvelwave}c) located with the red dot and dashed lines.}
    \label{fig:dispersionSED}
\end{figure}

To gain further insights into mode-conversion, we examine the vibrational spectra of both the periodic and aperiodic SL. The atomistic wave-packet simulation is approximately harmonic, making all scattering in the simulation elastic  \cite{schelling2002phonon,schelling2003multiscale,schelling2004kapitza}. For the periodic SL $\lambda_{1}$ phonon, plotting the mode-converted wavevector at the incident phonon frequency reveals that the mode-converted phonon exactly falls in the second Brillouin zone of the periodic SL dispersion relation (Fig.~\ref{fig:dispersionSED}a), evidencing the manifestation of coherent modes obeying the dispersion relation of the periodic SL through interference of incoherent phonons. 

In contrast to the periodic SL, the aperiodic SL lacks a well-defined phonon dispersion relation, rendering lattice dynamics unsuitable for its study.  One approach to explore its vibrational spectrum is computing the spectral energy density through equilibrium molecular dynamics simulations  \cite{thomas2010predicting,feng2015anharmonicity}. In Fig.~\ref{fig:dispersionSED}b, we present the spectral energy density of the aperiodic SL below the Debye temperature. The mode-converted phonon in the $\lambda_{3}$ case falls on a somewhat faint band in the second Brillouin zone. This finding reveals that, akin to the periodic SL, the considerable thermal conductivity observed in the aperiodic SL device stems from incoherent phonons converting to coherent modes defined by an approximate dispersion relation. Consequently, the overall level of mode-conversion appears subject to how the specific layering structure affects interference states and thus dispersion. 

In summary, through atomistic wave-packet simulations, we found a prominent transmission peak of incoherent phonons in the aperiodic SL; a surprising result considering the aperiodically arranged interfaces are expected to strongly scatter both particle-like and wave-like modes. The peak is owed to the mode-conversion of incoherent phonons to coherent modes defined by the approximate dispersion relation of the aperiodic SL. This behavior is similarly observed in the periodic SL where non-Bragg-reflected phonons can convert to coherent modes from the periodic SL dispersion relation and thus exhibit near-unity transmission. In both the periodic and aperiodic SL, the mode-converted phonons are found to lie outside the first Brillouin zone. Our results reveal that phonon transmission in both the periodic and aperiodic SL is dictated by the degree to which incoherent modes can convert to coherent modes defined by the device's band structure. From this, we infer that manipulating the band structure of the SL through changes in the layering pattern is a pathway to tailoring thermal conductivity to specific limits, depending on the application.

\section*{Supplementary Material}
See the supplementary material for Figs. S1a and Fig S1b.

\section*{Acknowledgements}
The authors thank the support from the Nuclear Regulatory Commission (award number: 31310021M0004) and the National Science Foundation (award number: 2047109). The authors also would like to acknowledge the support of Research and Innovation and the Cyberinfrastructure Team in the Office of Information Technology at the University of Nevada, Reno for facilitation and access to the Pronghorn High-Performance Computing Cluster.

\section*{Author Declarations}
\subsection{Conflict of Interest}
The authors have no conflicts to disclose.
\subsection{Author Contributions}
\textbf{Theodore Maranets:} Writing – review \& editing (equal), Writing – original draft (lead), Software (lead), Methodology (lead), Formal analysis (equal), Conceptualization (equal). \textbf{Yan Wang:} Writing – review \& editing (equal), Writing – original draft (equal), Supervision (lead), Funding acquisition (lead), Formal analysis (equal), Conceptualization (equal).

\section*{Data Availability}
The data that support the findings of this study are available from the corresponding author upon reasonable request.

\nocite{*}
\bibliography{references}% Produces the bibliography via BibTeX.

\end{document}